\documentclass[twocolumn,showpacs,prb,graphics,graphicx]{revtex4-1}
\usepackage{bm}
\usepackage[pdftex]{graphicx}
\usepackage{dcolumn}

\begin{document}

\title{Vortex-induced strain    and magnetization in type-II superconductors.}
\author{V. G. Kogan}
\affiliation{Ames Laboratory - DOE and Department of Physics, Iowa State University,
Ames, IA 50011}

\begin{abstract}
It is argued that the stress caused by vortex cores in the mixed state of superconductors may result in a field dependent contribution to the free energy and magnetization of measurable levels.   For sufficiently strong stress dependence of the critical temperature,  $\partial T_c/\partial p$, this contribution may result in the so-called ``second peak" in the field dependence of the {\it reversible} magnetization, the effect 
often masked by vortex pinning and  creep.
\end{abstract}


 \pacs{74.20.De,74.25.Ha,74.25.Wx,74.62.Fj}
\maketitle

The so-called second peak in the field dependence of the magnetization $M(H)$   in a number of type-II superconductors is  a long standing puzzle. The peak has been observed in magnetization loops  having a ``fish tail" shape so that the loop width of the irreversible magnetization increases with increasing field in  intermediate field range suggesting the critical current rise with increasing field in this domain.\cite{Larb}  

A few  explanations based on peculiarities of pinning and flux creep have been offered for this  apparently strange phenomenon.\cite{Cohen} These suggestions may well be correct but they do not cover all cases in which the second peak has been observed. Puzzling in particular is the fact that in some systems the second peak has been reported also in the {\it reversible} $M(H)$; the examples are NbSe$_2$, La$_{1.45}$Nd$_{0.40}$Sr$_{0.15}$CuO$_4$, and   CeCoIn$_5$.\cite{Sirenko,Finn,Almasan}

In this work, the second peak in reversible $M(H)$ is associated with the strain caused by  normal vortex cores embedded in the superconducting phase, a ``magneto-elastic" effect. The strains  arise  due to a small difference in densities of the normal and superconducting phases which is related to the stress dependence of the critical temperature $\partial T_c /\partial p$.\cite{LL2} It turned out recently that  this derivative in pnictides, and in Ca(Fe$_{1-x}$Co$_x$)$_2$As$_2$ in particular,\cite{dTc/dp} by one or two orders of magnitude exceeds   values for conventional superconductors making Fe-based pnictides especially favorable for observation of   magneto-elastic effects.



{ \bf   Strain caused by a single vortex}.  Consider vortex nucleation prior to 
which the superconductor has been strain free. We model 
the vortex core as a normal ($n$) cylinder of radius $\rho\sim\xi$, the coherence length,  
immersed in the superconducting ($s$) phase with a constant order 
parameter. This is a London-type approach\cite{KBMD} which suffices for qualitative estimates, although Ref.\,\onlinecite{Cano}   argues   that such an approach underestimates magneto-elastic effects.   

 Nucleation of the  normal core causes stress, since the $n$ phase 
has a larger specific volume $V_n$ as compared to $V_s$. The relative volume change $\zeta$ is related to the pressure dependence of the condensation energy or of the critical field $H_c$: \cite{LL2} 
\begin{equation}
\zeta = \frac{V_n-V_s}{V_s}=\frac{H_c}{4\pi}\frac{\partial H_c}{\partial 
p}\,.
\label{zeta}
\end{equation}

 The elastic  energy density in isotropic solids reads: \cite{LL}
\begin{equation}
F=\lambda u_{ll}^2/2 +\mu u_{ij}^2\,, 
\label{1}
\end{equation}
where $u_{ij}$ is the strain tensor  and $\lambda$, $\mu$ are 
${\rm Lam\acute{e}}$ coefficients; summation over double indices is implied. 
The stress tensor $\sigma_{ij} = \partial F/\partial 
u_{ij} = \lambda u_{ll}\delta_{ij} + 2\mu u_{ij}$, and the 
equilibrium condition $\partial \sigma_{ij}/\partial x_j \equiv  
\sigma_{ij,j} =0$ is given by  
\begin{equation}
\lambda u_{ll,i} +2\mu u_{ij,j}=0\,. 
\label{2}
\end{equation}
For brevity, the coma in   $u_{ik,j}$ is used to denote derivatives with respect to the coordinate $j$.
  
For a   single vortex  directed along $z$, the displacement ${\bf u}=(u_x,u_y,0)$ is radial in the plane $xy$, 
i.e., ${\rm curl}\,{\bf u}=0$ or ${\bf u}=\nabla \chi$, and 
$u_{\alpha \beta}=\chi_{,\alpha \beta}$ where $\chi$ is a scalar and $\alpha,\beta$ acquire only $x$ and $y$ values. The equilibrium condition (\ref{2}) reads $(\lambda 
  +2\mu) \chi_{,\alpha \beta \beta }=0 $ with the first integral 
\begin{equation}
\chi_{,\beta \beta } \equiv \nabla^2 \chi = C= {\rm constant}\,.
\label{chi}
\end{equation}
 To fix this constant, we note that $ \chi_{,\beta \beta }=u_{\beta \beta}$ describes compression and  is related the hydrostatic pressure within the system. For the problem of the strain caused by a single vortex in otherwise unrestrained crystal, the pressure is zero, and we have to solve $\nabla^2 \chi =0$ under the boundary condition   $u\to 0$ at large distances. Hence, the problem is the same as that of a linear charge in electrostatics: $\chi\propto \ln r$, $\bm r=(x,y)$. 
Hence,   we obtain:\cite{KBMD}
\begin{equation}
{\bf u}_s= \frac{\gamma_s \xi^2{\bf r}}{r^2}\,,\,\,\, 
u_{\alpha \beta}^{(s)}=\frac{\gamma_s\xi^2}{r^2}\left (\delta_{\alpha 
\beta}-\frac{2}{r^2}x_{\alpha}x_{\beta}\right );
\label{5}
\end{equation}
 where $\xi^2$ is introduced for convenience and the constant $\gamma_s$ is given below.   
 
At the core center    ${\bf u}(0)=0$;  we have   
\begin{equation}
{\bf u}_n=-\gamma_n {\bf r},\,\,\, u_{\alpha\beta}^{(n)}=-\gamma_n 
\delta_{\alpha\beta}
\label{4}
\end{equation}
in the core interior. The constants $\gamma$'s  are evaluated by using  boundary conditions at the interface: \cite{KBMD} 
 \begin{equation}
\gamma_n = \frac{\zeta\mu}{2(\lambda+2\mu)}\,,\,\,\, 
\gamma_s = \frac{\zeta(\lambda+\mu)}{2(\lambda+2\mu)}\,. 
\label{7}
\end{equation}
 
The displacement ${\bm u}_s$ of Eq.\,(\ref{5}) is analogous to the electric field of a charge  with linear  density $\gamma_s\xi^2/2$ situated at the origin. Hence, the vortex can be considered as the linear source  of deformation $\bm u$ outside the core,  whereas the scalar potential $\chi$ satisfies
 \begin{equation}
\nabla^2\chi = 2\pi \gamma_s\xi^2\delta(\bm r)\,.
\label{7}
\end{equation}


{\bf   Vortex lattice.}
Consider now a 2D periodic lattice of vortices at positions $\bm a$ in an infinite sample.    At first sight, the potential $\chi$ should obey
 \begin{equation}
\nabla^2\chi = 2\pi \gamma_s\xi^2\sum_{\bm a}\delta(\bm {r-a})\,. 
\label{8}
\end{equation}
The electrostatic analogy, however, shows that this equation cannot have a bound solution, whereas we are interested in periodic $\chi(\bm  r )$  to describe the infinite vortex lattice. We, therefore, introduce a background ``charge density" of a sign opposite to $\gamma_s\xi^2/2$ to make the system ``quasi-neutral". In other words, the condition for a periodic  $\chi$ to exist is
 \begin{equation}
\int d{\bm r}\left( 2\pi \gamma_s\xi^2\sum_{\bm a}\delta(\bm {r-a})+C\right)=0\,. 
\label{9}
\end{equation}
This translates to $2\pi \gamma_s\xi^2N+ CA=0$ where $N$ is the total number of vortices and $A$ is the area of the sample crossection perpendicular to the induction $\bm B$; $N/A=B/\phi_0$ is the density of vortices. Hence, $C = - 2\pi \gamma_s\xi^2B/\phi_0$ 
 and we have to look for solutions of
 \begin{equation}
\nabla^2\chi = 2\pi \gamma_s\xi^2\left[\sum_{\bm a}\delta(\bm {r-a})-\frac{B}{\phi_0}\right]\,, 
\label{11}
\end{equation}
an equation consistent with the equilibrium condition $ \nabla^2 \chi  =$ constant. 

The general solution of this equation was discussed in Ref.\,\onlinecite{K75}. Dealing with periodic solutions, one can consider $\chi(\bm r)$ in a single cell under the   condition $\partial\chi/\partial {\bm n}=0$ at the cell boundary  (${\bm n}$ is the normal to the  boundary). The potential within a cell centered at ${\bm a}=0$  satisfies 
 \begin{equation}
\nabla^2\chi = 2\pi \gamma_s\xi^2\left[ \delta(\bm {r})-\frac{B}{\phi_0}\right]\,. 
\label{12}
\end{equation}

The form of the unit cell depends on the vortex lattice structure which is hexagonal (triangular) in isotropic case of interest here. For this lattice, the boundary is a hexagon which - in the Wigner-Zeitz   approximation - can be replaced with a circle. The cylindrically symmetric solution satisfying   $\chi^\prime(R)=0$ with $\pi R^2=\phi_0/B$ is
 \begin{equation}
 \chi = \gamma_s\xi^2\left(\ln \frac{r}{r_0}  -\frac{\pi B}{2\phi_0}\,r^2\right)\,; 
\label{13}
\end{equation}
$r_0$ is an arbitrary constant irrelevant for the following. 

The crystal displacement has only one component:
\begin{equation}
 u_r = \gamma_s\xi^2\left(  \frac{1}{r }  -\frac{\pi B}{ \phi_0}\,r \right)\,.
\label{14}
\end{equation}
The strain tensor in cylindrical coordinates\cite{LL} has two non-zero components:
\begin{eqnarray}
 u_{rr} &=&\frac{\partial u_r}{\partial r}= -\gamma_s\xi^2\left(  \frac{1}{r^2 } +\frac{\pi B}{ \phi_0}  \right)\,,\nonumber\\
  u_{\varphi\varphi} &=&\frac{  u_r}{  r}=  \gamma_s\xi^2\left(  \frac{1}{r^2 } -\frac{\pi B}{ \phi_0}  \right)\,.
\label{15}
\end{eqnarray}
 
The   elastic energy density averaged over the cell is
\begin{eqnarray}
  F_{el} =   \frac{B}{\phi_0}\int_{\rho}^R 2\pi r\, dr  \left[ \lambda u_{\alpha\alpha}^2(  r)/2 +\mu u_{\alpha\beta}^2(  r)  \right],
 \label{eq16}
\end{eqnarray}
where the lower integration limit is the core radius on the order of $\xi$.
Within the London approach   one cannot determine the radius $\rho$; we will choose it below as to have the elastic contribution to magnetization to be zero at the upper critical field $H_{c2}$. 
 
A straightforward evaluation gives:
\begin{eqnarray}
 F_{el} &=&  \frac{\tilde{\lambda}}{2} \gamma_s^2 b^2 \left(1-\frac{\rho^2}{2\xi^2}\,b\right)\,,\qquad b = \frac{B}{H_{c2} }\,,   \label{eq17}
\end{eqnarray}
and 
\begin{eqnarray}
 \tilde{\lambda}=   \lambda + \mu  +\mu\frac{2\xi^2}{\rho^2b} \,   
 \label{eq18}
\end{eqnarray}
is a quantity on the order of the elastic constants.  

 {\bf   Parameter $\bm   \zeta$ in terms of $\bm{ \partial T_c /\partial p $}.} The stress dependence of the condensation energy $\partial (H_c^2/8\pi)/\partial p$, to which the coefficient $\gamma_s$ is proportional, can be evaluated only within a detailed microscopic theory to  account for evolution of the band structure and of the coupling responsible for superconductivity with pressure. Such a calculation, if possible, would be material specific. Instead, we resort to a qualitative  approach   to see how the the vortex induced strain could affect macroscopic   properties of type-II superconductors. 
 
First, the derivative in Eq.\,(\ref{zeta}) can be expressed in terms of the  measured $\partial T_c /\partial p$:
\begin{equation}
  \frac{\partial (H_c^2/8\pi)}{\partial p}= \frac{ \partial (H_c^2/8\pi)}{\partial  T_c} 
  \frac{\partial T_c  }{\partial p} \,.  \label{derivative}
\end{equation}
Unfortunately, there is no  simple enough expression for the condensation energy   $H_c^2/8\pi=F_n-F_s $ for arbitrary temperatures, fields, and scattering regimes. 
The exception is  the case of a strong pair-breaking considered originally by Abrikosov and Gor'kov,\cite{AG} who argued that due to extra suppression of the order parameter by, e.g., spin-flip scattering, the GL energy expansion holds for all temperatures down to $0$. This argument has recently been specified for  materials with zero order parameter average (like d-wave or $\pm s$ for iron-based pnictides).\cite{pair-break} This scheme will be used below mostly because its formal simplicity, although the qualitative results obtained have a broader applicability. 
 
The zero-field condensation energy  for gapless state is 
\begin{equation}
  \frac{ H_c^2}{8\pi }= A(T_c^2-T^2)^2 \,,\quad A \sim \frac{N(0)\tau_+^2}{\hbar^2}
   \label{pairbr}
\end{equation}
where  $N(0)$ is the density of states, and $1/\tau_+=1/\tau +1/\tau_m$ whereas $1/\tau $ and $1/\tau_m $ are the transport and pair-breaking scattering rates.  One then finds:
\begin{equation}
  \frac{\partial  H_c^2 /8\pi  }{\partial T_c}=  \frac{  H_{c0}^2 }{2\pi T_c}(1-t^2) \,,\qquad t=T/T_c\,.  
   \label{derivative1}
\end{equation}
Thus, we estimate: 
\begin{equation}
\zeta \approx   \frac{ H_{c0}^2  }{2\pi T_c}(1-t^2) \frac{\partial T_c}{\partial p}\,.  
\label{zeta2}
\end{equation}
 
Also, within the gapless state, the upper critical field and the London penetration depth have simple $T$ dependencies used below:
\begin{equation}
  H_{c2} = H_{c2,0}(1-t^2)  \,,\quad \lambda_{L }^2=\lambda_{L,0}^2/  (1-t^2)\,.  
  \label{pairbr-a}
\end{equation} 

{\bf Magnetization}. The free energy density of   the mixed state is 
\begin{equation}
F =F_0+B^2/8\pi + F_L +F_{el} \,,
\label{energy2}
\end{equation}
where $F_0$ is the free energy of the uniform state in zero field. 
The London energy of the vortex lattice in intermediate fields is given by
\begin{equation}
 F_L \approx \frac{\phi_0B}{32\pi^2\lambda_L^2}\,\ln\frac{\eta H_{c2}}{B} \, 
\label{energyL}
\end{equation}
with $\eta\sim 1$. The elastic part is obtained with the help of Eqs.\,(\ref{eq17}), (\ref{zeta2}), and (\ref{pairbr-a}):
 \begin{eqnarray}
 F_{el}   \approx      \overline{\lambda } \left[\frac{ H_{c0}^2 }{2\pi T_c}\frac{\partial T_c}{\partial p} \frac{B }{H_{c2,0} } \right]^2 \left(1-\frac{\rho^2B}{2\xi^2H_{c2}}\right) . 
\label{Fel}
\end{eqnarray}
  Here, $\overline{\lambda}\sim 10^{12}\,$erg/cm$^3$    is a new combination  of $ \lambda $ and $\mu$. 
  
  Both $F_L$ and $F_{el}$ are evaluated here within London   approach for  $H_{c1}\ll B\ll H_{c2}$ and, therefore, fail near both $H_{c2}$ and $H_{c1}$. Still, we can force the magnetization  to be zero at $H_{c2}$ by  setting $\eta=e\approx 2.718$ and $\rho=2\xi/\sqrt{3}$. We then obtain:  
\begin{eqnarray}
 &M&  = \frac{B-H}{4\pi}= \frac{B}{4\pi} -\frac{\partial F}{\partial B} =M_L+M_{el}\,, \label{27}\\
&M_L& =- \frac{\phi_0 }{32\pi^2\lambda_L^2}\,\ln\frac{H_{c2}}{B}\,, \label{28}\end{eqnarray}
where $M_L$ is the standard London part. 
The average penetration length $\lambda_L$, that  governs the  field distribution in  the mixed state,  depends on the field because the average order parameter $\Delta$  is suppressed by the field. This dependence is relevant in particular 
  near $H_{c2}$, since there the averaged order parameter $\overline {\Delta^2} \propto (1-B/H_{c2})$ which translates to $\lambda\propto1/\sqrt{1-B/H_{c2})}$.\,\cite{clem}  We take this field dependence to characterize qualitatively $M_L(T,B)$. 

The elastic contribution to $M$ is
 \begin{eqnarray}
M_{el}&=&-2\overline{\lambda }  \left[\frac{ H_{c0}^2 }{2\pi T_cH_{c2,0}}\frac{\partial T_c}{\partial p} \right]^2B\left(1-\frac{B}{H_{c2}}\right).\qquad 
 \label{29}
\end{eqnarray}
It is worth noting that this contribution   is diamagnetic  
and has a minimum at $B=H_{c2}/2$ in a reasonable agreement with  data of Refs.\,\onlinecite{Finn} and \onlinecite{Almasan}. 

\begin{figure}[h ]
\includegraphics[width=7cm]{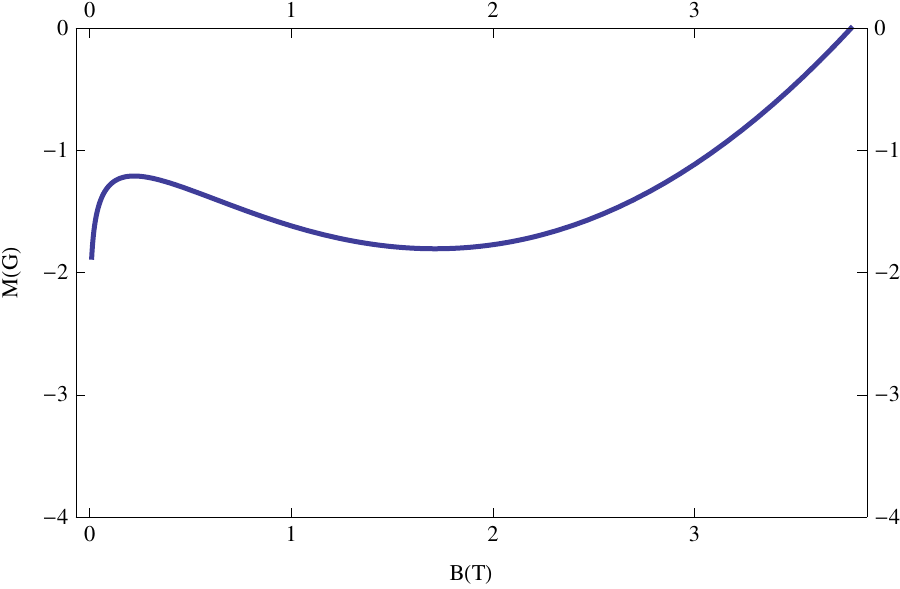}
 \caption{(Color online)  The magnetization $M$ {\it versus} applied field $ B $ according to Eqs.\,(\ref{27})--(\ref{29})  for parameters given in the text. }
 \label{fig1}
 \end{figure}

 Taking $\overline\lambda\approx 10^{12}\,$erg/cm$^3$, $H_{c2,0}=5.9\,$T, $H_{c0}=0.35\,$T, $\lambda_L(0)\approx 3.5\times 10^{-5}\,$cm, $T_c\approx 10.5\,$K, and $\partial T_c/\partial p\approx 3 \times 10^{-9}\,$K cm$^3$/erg=30\,$ \,$K/GPa  one obtains   $M(B)$   shown in Fig.\,\ref{fig1}. These numbers roughly correspond to  parameters for 
 La$_{1.45}$Nd$_{0.40}$Sr$_{0.15}$CuO$_4$.\cite{Finn}
 It should be noted that $\partial T_c/\partial p$ for this particular material was not measured, but the data on a similar crystal, 
La$_{1.44}$Nd$_{0.40}$Sr$_{0.14}$CuO$_4$,\cite{Hucker} show that  $\partial T_c/\partial p|_{p\to 0}$ exceeds 15\,K/GPa. 

\begin{figure}[h ]
\includegraphics[width=7.5cm]{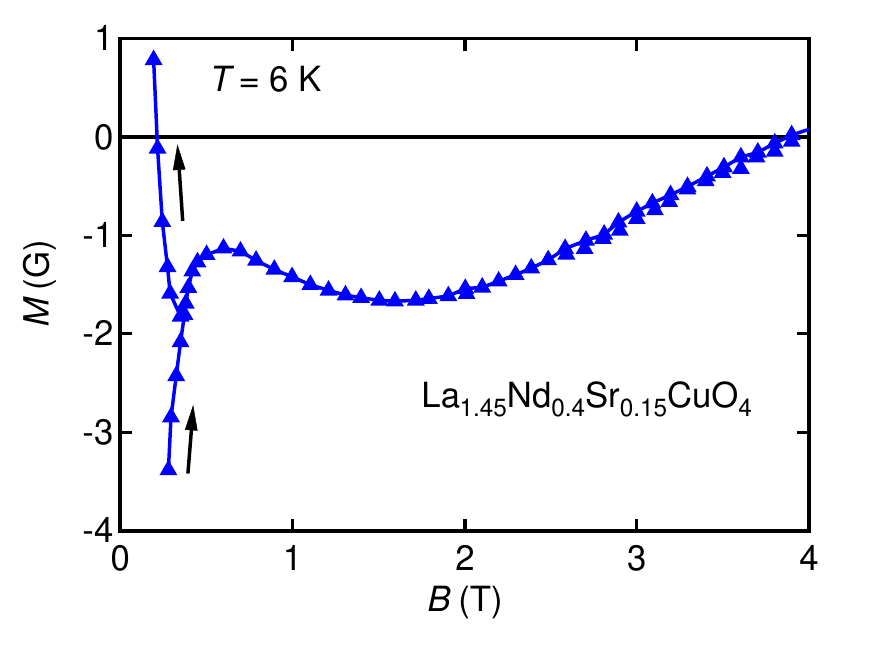}
 \caption{(Color online)  The magnetization $M(B)$ measured in increasing and decreasing fields along the $c$ crystal axis; the data are compiled from Ref.\,\onlinecite{Finn}. It is seen that   $M(B)$ is reversible for $B> 0.4\,$T. }
 \label{fig2}
 \end{figure}
 
 Qualitatively, the calculated $M(B)$ is similar to that recorded by Ostenson et al \cite{Finn} and shown in Fig.\,\ref{fig2}. The major features of the data are reproduced by the model remarkably well. One must bear in mind that the London-type isotropic model cannot pretend for quantitative agreement with data near  $H_{c2}$ and for $B\to 0$. Besides, in anisotropic materials one cannot use the data on the $T_c$ dependence of hydrostatic pressure $p$ as a fair representative of actual dependence of  $T_c$ on the stress in the plane perpendicular to the applied field. Moreover, the use of  temperature dependencies of quantities involved characteristic of the gapless situation was only justified by formal simplicity of derivations. Nevertheless, one   concludes  that the evidence for the vortex induced strain as responsible for non-monotonic behavior of $M(B)$ in materials with a large $\partial T_c/\partial p $ is quite strong. \\

{\bf Discussion.} The elastic contribution to the vortex-vortex interactions was studied in a number of publications, see Refs.\,\onlinecite{KBMD}, \onlinecite{Cano} and references therein. It has been shown that this contribution is responsible for the observed flux-line lattice structures in fields tilted with respect to the $c$ axis of NbSe$_2$. The observed structures  cannot be explained by   London interactions alone; the structures seen in experiments in fact correspond to the maximum of the London energy.  Vortex lattices are extremely sensitive to a number of factors, among which the nonlocal corrections to London interactions were proven to be important.\cite{cubVL,boroVL} Energy differences between various vortex lattices are exceedingly small.  
 It is shown in this work that the elastic deformations caused by vortices may influence such a quantity as magnetization involving much larger energies. 

The model suggested here is profoundly qualitative. Materials to which the model is   applied are anisotropic for which one needs to know a number of   elastic constants. We lumped all this complexity to one number, $\overline\lambda\approx 10^{12}\,$erg/cm$^3$. 
Our estimates of material parameters and in particular of $\partial T_c/\partial p $   needed for evaluation of elastic contribution to magnetization are quite crude. In particular, it is hard to get reliable values of $\partial T_c/\partial p $ for $p\to 0$ since commonly people are interested in high pressures.\cite{Hucker}  

We model vortices as having normal cores surrounded by superconductor with unperturbed order parameter so that the condensation energy  is just $H_c^2/8\pi$ which is so only far from $H_{c2}$, see also Ref.\,\onlinecite{Cano}. Hence, the model fails in high fields approaching $H_{c2}$. We have used simplified $T$ dependencies of $H_{c2}$ and $\lambda_L$ corresponding to the strong pair-breaking  situation. Still, all these uncertainties notwithstanding, the model reproduces qualitatively the behavior of $M(B)$ with the second peak.  

The interpretation of the second peak in $M(H)$ as an equilibrium thermodynamic property of deformable type-II superconductors is new; it  differs from traditional models based on various  defects-related irreversible material properties. The latter are always present, of course, and make it difficult to extract relatively weak magneto-elastic properties of vortex lattices. It should be noted, however, that well-pronounced 2nd peaks in reversible $M(H)$ must result in similar peaks in irreversible magnetization loops, in other words, the fish-tail loops seen in layered materials may also be caused by the vortex induced strain. 

Layered materials having strong stress dependencies of $T_c$ are therefore good candidates not only for unmasking equilibrium magneto-elastic phenomena from the background of strong irreversibilities, but for understanding the irreversible fish-tails as well. Recently, the pressure dependence of $T_c$ in  Ca(Fe$_{1-x}$Co$_x$)$_2$As$_2$ was found to reach $\partial T_c/\partial p\approx -60\,$ K/GPa, which is by a factor of 100 more than  in ``conventional" superconductors. 
  The present work  suggests that   magneto-elastic effects should be studied in materials with large  $\partial T_c/\partial p $ with an emphasis on macroscopic magnetization, the problem deserving more experimental and theoretical attention.   \\

 The author is grateful to S. Bud'ko, P. Canfield, R.~Prozorov, J. Clem, D. Finnemore, M. Tanatar, V.~Taufour and other colleagues  for many illuminating discussions. The Ames Laboratory is supported by the Department of Energy, Office of  Basic Energy Sciences, Division of Materials Sciences and Engineering under Contract No. DE-AC02-07CH11358.

\end{document}